\documentclass[twocolumn]{revtex4}
\usepackage{graphicx}

\begin{document}
\def\ket{\rangle}
\def\bra{\langle}
\def\w{\omega}
\def\W{\Omega}
\def\lr{\leftrightarrow}
\def\ud{\updownarrow}
\title{Two-step Quantum Key Distribution Schemes Using Polarization and Frequency
Doubly Entangled Photons}
\author{Chuan Wang}
\affiliation{Department of Physics, Tsinghua University, Beijing,
100084, People's Republic of China}
\author{Li Xiao}
\affiliation{Tsinghua National Laboratory For Information Science
and Technology, Tsinghua University\\
Department of Electronic Engineering, Tsinghua University, Beijing,
100084, People's Republic of China}
\author{Wan-ying Wang}
\affiliation{Department of Physics, Tsinghua University, Beijing,
100084, People's Republic of China}
\author{Gui Lu Long}
\affiliation{Key Laboratory of Atomic and Molecular Sciences and
Department of Physics, Tsinghua University, Tsinghua National
Laboratory For Information Science and Technology, Tsinghua
University, Beijing, 100084, People's Republic of China}

\date{\today }

\begin{abstract}
A two-step quantum key distribution protocol using frequency and
polarization doubly entangled photons is proposed. In this protocol,
information is encoded by a unitary operation on each of the two
doubly entangled photons and sent from Alice to Bob in two steps.
State measurement device is designed.  The security of the
communication is  analyzed.
\end{abstract}

 \maketitle

Quantum communication, especially quantum key distribution (QKD) is
an important branch of quantum information\cite{nielsen} which
provides a secure way for creating secret keys between the
communication parties called Alice and Bob. In the past few years,
quantum key distribution has progressed
quickly\cite{bb84,ekert91,bbm92,b92,six,gisin,guo,lo,core,hwang,mazhao,wangx,qsdc}
since Bennett and Brassard presented the BB84 QKD protocol
\cite{bb84}. Some QKD protocols are based on the dense coding using
Bell states \cite{bw}. One such protocol is the two-step protocol in
which the communication is realized in two steps \cite{liu,deng}, it
can not only be used for QKD, but also for quantum secure direct
communication in which information is transmitted directly without
using another classical communication of the ciphered text
\cite{deng,cai,yan,zhangzj,wang,qsdc}. Meanwhile experiments on
quantum dense coding based on entangled photon pairs and continuous
variables have also been reported in \cite{mattle,kimble,zhang}.

Entangled photon pairs have become a very important resource  for
quantum communication. Experimentally it is usually produced by
using spontaneous parametric down conversion (SPDC) \cite{kwiat}, in
which a nonlinear crystal pumped by lasers to create two photons
marked with idler (i) and signal (s) respectively under the phase
match conditions: $\omega_{p}=\omega_{s}+\omega_{i}$,
$k_{p}=k_{s}+k_{i}$. These entangled photons are usually the
polarization entangled photons. Recently, Ravaro et al have
experimentally demonstrated the generation of  frequency and
polarization doubly entangled photons (DEPs) using AlGaAs
multi-layer waveguide \cite{ravaro}. The high conversion efficiency
of the DEP pair has attracted much attention in recent years. It is
interesting to study the quantum communication protocols based on
them. In this paper, we present a secure communication scheme based
upon the polarization and frequency DEPs. Because the DEP pairs
possess an additional degree of freedom in frequency,  this protocol
also has the advantage of a higher capacity of coding.

DEP and its application was first introduced by Aolita and Walborn
\cite{aolita}. Using nonlinear AlGaAs waveguide, Ravaro et al.
proposed a scheme theoretically of generating two counter
propagating photons in the state:
\begin{eqnarray}
|ent\rangle=\frac{\eta_{1}}{\sqrt{|\eta_{1}|^{2}+|\eta_{2}|^{2}}}
|{\omega}_{s},H\rangle|{\omega}_{i},V\rangle \nonumber\\
+\frac{\eta_{2}}{\sqrt{|\eta_{1}|^{2}+|\eta_{2}|^{2}}}
|{\omega}_{s'},V\rangle|{\omega}_{i'},H\rangle,
\end{eqnarray}
where $|H\ket$ $(|V\ket)$ represents the horizontal polarization of
the electric(magnetic) field. If $\eta_{1}=\eta_{2}$, the state
becomes frequency and polarization maximally double entangled state.

Now we focus on the maximally entangled doubly entangled states.
First lets inspect the following state
\begin{equation}
\Psi^{+}_{ab}=\frac{1}{\sqrt{2}}(|H,\omega_{s}\rangle|V,\omega_{i}\rangle
+|V,\omega_{s'}\rangle|H,\omega_{i'}\rangle),
\end{equation}
as said before, $H$ and $V$ are polarizations of photons,
$\omega_{s}$, $\omega_{i}$, $\omega_{s'}$, $\omega_{i'}$ are the
frequencies of the signal and idler photons. Instead of producing
entangled photons with fixed frequencies in the polarization
(singly) entangled photon state, doubly entangled photon pairs have
two possible frequency combinations, namely it can either in state
with frequencies $\omega_s$  and $\omega_i$, or with frequencies
$\omega_{s'}$  and $\omega_{i'}$.  Photon $a$ is either in the state
$|H,\omega_{s}\rangle$ or in $|V,\omega_{s'}\rangle$, photon $b$ is
either in the state $|H,\omega_{i}\rangle$ or in
$|V,\omega_{i'}\rangle$.

Local unitary operations acting on the two particles transforms the
state $\Psi^{+}_{ab}$ to other eight bases  shown below
\begin{eqnarray}
\Phi^{\pm}_{ab}=\frac{1}{\sqrt{2}}(|H,\omega_{s}\rangle|H,\omega_{i}\rangle
\pm|V,\omega_{s'}\rangle|V,\omega_{i'}\rangle);\\
\Psi^{\pm}_{ab}=\frac{1}{\sqrt{2}}(|H,\omega_{s}\rangle|V,\omega_{i}\rangle
\pm|V,\omega_{s'}\rangle|H,\omega_{i'}\rangle);\\
\Gamma^{\pm}_{ab}=\frac{1}{\sqrt{2}}(|V,\omega_{s}\rangle|H,\omega_{i}\rangle
\pm|H,\omega_{s'}\rangle|V,\omega_{i'}\rangle);\\
\Upsilon^{\pm}_{ab}=\frac{1}{\sqrt{2}}(|V,\omega_{s}\rangle|V,\omega_{i}\rangle
\pm|H,\omega_{s'}\rangle|H,\omega_{i'}\rangle).
\end{eqnarray}

The correspondence between unitary operation and state is given  in
Table.\ref{t1}. As listed in the table, $U_{a}\otimes U_{b}$
operation transforms state $\Psi^{+}_{ab}$ to a unique state in the
basis set, and Alice and Bob agree that each state corresponds to a
3-bit code word given in the table.

The two-step QKD protocol based on DEPs is described in detail as
follows.

\noindent Two-step Double-QKD protocol

(Step 1): Alice,  the sender, prepares a series of DEP pairs in
$\Psi^{+}_{a,b}$. For each state, She makes a unitary operation on
photon $b$. The operations are chosen from the group
$\{I,\sigma_{x},\sigma_{z},i\sigma_{y}\}$. This transforms state
$\Psi^{+}_{a,b}$ into one of  the state in
$\{\Phi^{\pm}_{a,b},\Psi^{\pm}_{a,b}\}$. She then sends the
$b$-photon sequence
 $[P_{1}(b),P_{2}(b),\cdots P_{n}(b)]$ to Bob.

(Step 2): Bob receives and saves the photons. Then they perform a
security check, the strategy will be discussed later.

(Step 3): Alice and Bob check Eve by public comparison. If the error
rate is lower than a security bound, they confirm that the first
round communication is secure.

(Step 4): Alice encodes key information on the $a$ photons using
$\{I,\sigma_{x},\sigma_{z},i\sigma_{y}\}$ operations. After that,
she sends the $a$ photons sequence $[P_{1}(a),P_{2}(a),\cdots
P_{n}(a)]$ to Bob.

(Step 5):Bob receives the photons and makes joint  measurement to
distinguish the eight different states.

\begin{figure}
\begin{center}
\includegraphics[width=9cm,angle=0]{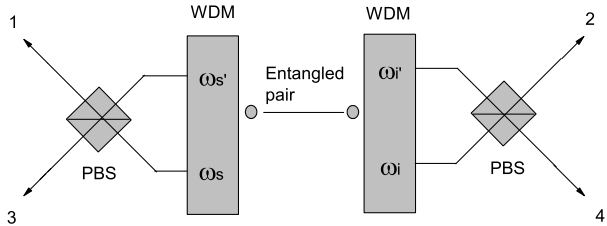}
\caption{ Bell state measurement device. \label{f1}}
\end{center}
\end{figure}

Next we focus on the devices for state  measurement.
 The measurement device is shown in Fig.\ref{f1}. When a DEP enters
 Bob's measurement device, one to the left wavelength-division multiplexing device(WDM) and
the other to the right,
  photons with different frequencies can be distinguished. The two photons leave the port of
each WDM in either the $\omega_{s}/\omega_i$ port or the
$\omega_{s'}/\omega_{i'}$ port.

\begin{figure}
\begin{center}
\includegraphics[width=8cm,angle=0]{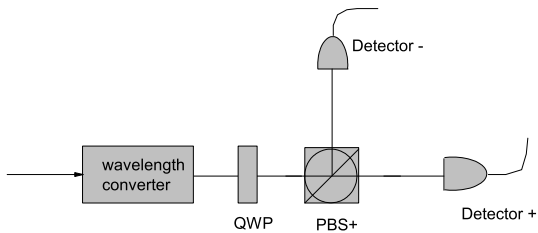}
\caption{ $\hat{X}$ bases measurement device. Here QWP is the
quarter wave plate and PBS+ is the $\sigma_{x}$ basis polarizing
beam splitter. \label{f2}}
\end{center}
\end{figure}

Then, each photon passes through a polarizing beam splitter and
enters a device shown in Fig.\ref{f2}  being detected by one of the
two detectors at each port. Table \ref{t2} gives the correspondence
between the ports at which detectors triggered  and the states. Here
there is still a two-fold degeneracy, and it is further
distinguished by the two detectors at each port.

At each port, we place a wavelength converter and make a
$\sigma_{x}$ measurement as  shown in Fig.\ref{f2}.

In the measurement for example, at time $t$ the detectors on port 1
and port 4 triggers, Bob knows the state are $\Psi^{\pm}$, triggers
on port 3, 2 indicates $\Gamma^{\pm}$ and so on. Then passing the
two photons to wavelength converter \cite{yoo}. Then the state
degenerates to Bell state:
$\Psi^{\pm}_{Bell}=\frac{1}{\sqrt{2}}(|HV\rangle \pm |VH\rangle)$.
Following a quarter wave plate Bob transforms the $H,V$ bases to
$|+x\rangle=|H\rangle+|V\rangle,|-x\rangle=|H\rangle-|V\rangle$
bases. In this way, the states $\Psi^{\pm}$ can be transformed to
$|+x\rangle|+x\rangle-|-x\rangle|-x\rangle$ and
$|+x\rangle|-x\rangle-|-x\rangle|+x\rangle$. Bob performs a
$\sigma_{x}\otimes\sigma_{x}$ measurement and distinguishes the
state completely since the polarizations are either parallel or
antiparallel. The other three bases can also be distinguished. In
this way, the eight states can be completely distinguished.

There are two security checking strategies in our scheme, one is
``decoy state" strategy and the other is wavelength convertor
strategy.

The ``decoy state" scheme uses single photons. When the
communication starts, Alice inserts single photons with frequency
$\omega_{i}$ or $\omega_{i'}$ and polarization state $|H\rangle$,
$|V\rangle$, $|H+V\rangle$, $|H-V\rangle$ randomly in the $b$ photon
series as ``decoy state" photons. After Bob receives the photons,
Alice announces the positions of the single photons publicly. Bob
chooses to measure them either in the $\sigma_{z}$ or the
$\sigma_{x}$ bases. Bob then declares the bases and Alice checks
which bases they match. If they happen to choose the same bases,
their results are in agreement. If there are eavesdropping, there
will be error. Thus after comparison, if the error rate is lower
than a security bound, they confirm that the communication is
secure. Otherwise they drop the data and restart the communication.

In the wavelength convertor strategy, after Bob receives $b$ photons
from Alice, he chooses some photons randomly and performs a
wavelength conversion. He then announces the positions of photons to
Alice. Alice performs the same operation. The wavelength of
$\omega_{s}$, $\omega_{s'}$ and $\omega_{i},\omega_{i'}$ turn to
$\omega_{s_{0}}=min(\omega_{s}, \omega_{s'})$, $\omega_{i_{0}}=
min(\omega_{i},\omega_{i'})$ respectively. In this way, the DEPs
degenerate to Bell states:
\begin{eqnarray}
|\Phi^{\pm}_{Bell}\rangle=\frac{1}{\sqrt{2}}(|HH\rangle
\pm|VV\rangle);\nonumber\\
|\Psi^{\pm}_{Bell}\rangle=\frac{1}{\sqrt{2}}(|HV\rangle
\pm|VH\rangle).
\end{eqnarray}
They both choose to measure the photons in either $\sigma_{z}$ or
$\sigma_{x}$ bases, and announce the bases so as to pick up the
matched particles. Finally they compare the results and check the
error rate to confirm the security.

In the following we will analyze the security of the protocol under
the intercept-resend(IR) attack. Eve hides herself in the
communication channel and intercepts the "$b$" photons and measures
them, and then prepares a series of photons according to her
measurement, and resends them to Bob. Because the interception
destroys the entanglement between the $a$ and $b$ photons, Alice and
Bob may discover the eavesdropping in the second security checking
method. Also this IR attack will introduce 25\% of errors on the
``decoy state" photons, so it is easily to be discovered by the
first security checking method.

In conclusion, we have proposed a two-step QKD protocol using DEPs.
The measurement device in distinguishing the eight states is
designed. We also made an analysis on its security. Since the DEPs
has a higher generation efficiency, it maybe an alternative
entanglement source for communication.

This work is supported by the National Fundamental Research Program
Grant No. 2006CB921106, China National Natural Science Foundation
Grant Nos. 10325521, 60433050.

\begin{table}
 \caption{The correspondence between unitary operations and
states}\label{t1}
\begin{center}
\begin{tabular}{cccccc}
  \hline
  key & operation & state & key & operation & state \\
   \hline
  000 & $I\otimes I$ & $\Psi^{+}$ & 100 & $\sigma_{x}\otimes I$ & $\Upsilon^{+}$ \\
 \hline
  001 & $I\otimes \sigma_{z}$ & $\Psi^{-}$ & 101 & $\sigma_{x}\otimes\sigma_{z}$ & $\Upsilon^{-}$ \\
 \hline
  010 & $I\otimes \sigma_{x}$ & $\Phi^{+}$ & 110 & $\sigma_{x}\otimes\sigma_{x}$ & $\Gamma^{+}$ \\
 \hline
  011 & $I\otimes i\sigma_{y}$ & $\Phi^{-}$ & 111 & $\sigma_{x}\otimes i\sigma_{y}$ & $\Gamma^{-}$ \\
 \hline
  000 & $\sigma_{z}\otimes I$ &$\Psi^{-}$ & 100 & $i\sigma_{y}\otimes I$ & $\Upsilon^{-}$ \\
 \hline
  001 & $\sigma_{z}\otimes \sigma_{z}$ & $\Psi^{+}$  & 101 & $i\sigma_{y}\otimes \sigma_{z}$ & $\Upsilon^{+}$ \\
 \hline
  010 & $\sigma_{z}\otimes \sigma_{x}$ & $\Phi^{-}$ & 110 & $i\sigma_{y}\otimes \sigma_{x}$ & $\Gamma^{-}$ \\
 \hline
  011 & $\sigma_{z}\otimes i\sigma_{y}$ & $\Phi^{+}$ & 111 & $i\sigma_{y}\otimes i\sigma_{y}$ & $\Gamma^{+}$ \\
  \hline
\end{tabular}
\end{center}

\end{table}

\begin{table}
\caption{The correspondence between ports in Bell state measurement device and
states}\label{t2}
\begin{center}
\begin{tabular}{ccccccc}
  \hline
  Triggered Port && Corresponding states \\
  \hline
  1,2 &&  $\Phi^{\pm}$ \\
  \hline
  1,4 &&  $\Psi^{\pm}$ \\
  \hline
  3,2 &&  $\Gamma^{\pm}$ \\
  \hline
  3,4 &&  $\Upsilon^{\pm}$ \\
  \hline
\end{tabular}
\end{center}

\end{table}

\end{document}